\documentclass[prb,fleqn,12pt,showpacs]{revtex4}
\usepackage{epsfig}
\usepackage{graphicx}
\usepackage{amsfonts}
\begin{document}
\title{Multi-orbital quantum antiferromagnetism in iron pnictides --- 
effective spin couplings and quantum corrections to sublattice magnetization}
\author{Sayandip Ghosh, Nimisha Raghuvanshi, Shubhajyoti Mohapatra, Ashish Kumar, and Avinash Singh}
\address{Department of Physics, Indian Institute of Technology Kanpur 208016, India} 
\email{avinas@iitk.ac.in}
\date{\today} 

\begin{abstract}
Towards understanding the multi-orbital quantum antiferromagnetism in iron pnictides, effective spin couplings and spin fluctuation induced quantum corrections to sublattice magnetization are obtained in the $(\pi,0)$ AF state of a realistic three band interacting electron model involving $xz$, $yz$, and $xy$ Fe 3d orbitals. The $xy$ orbital is found to be mainly responsible for the generation of strong ferromagnetic spin coupling in the $b$ direction, which is critically important to fully account for the spin wave dispersion as measured in inelastic neutron scattering experiments. The ferromagnetic spin coupling is strongly suppressed as the $xy$ band approaches half filling, and is ascribed to particle-hole exchange in the partially filled $xy$ band. The strongest AF spin coupling in the $a$ direction is found to be in the orbital off diagonal sector involving the $xz$ and $xy$ orbitals. First order quantum corrections to sublattice magnetization are evaluated for the three orbitals, and yield a significant $37\%$ average reduction from the Hartree-Fock value. 
\end{abstract}
\maketitle
\newpage

\section{Introduction}
The rich phase diagram exhibited by iron pnictides 
\cite{Zhao2008nature,Nandi2010} including magnetic, structural, and superconducting phase transitions \cite{Fernandes2014} have stimulated intensive investigations aimed at detailed understanding of their macroscopic physical behavior in terms of their complex multi-orbital electronic structure as revealed by first-principle calculations \cite{Mazin2008,Haule2008,Singh2008,Nekrasov2008,Zhang2009,Graser2009,Ikeda2010,Graser2010} and angle resolved photoemission spectroscopy (ARPES) experiments \cite{Yi2009,Kondo2010,Yi2011,Brouet2012,Kordyuk2013}. The magnetic state exhibits
$(\pi,0)$ magnetic ordering of Fe moments in the $a-b$ plane, with a concomitant structural distortion $a>b$, possibly correlated with the ferro orbital order $n_{xz} > n_{yz}$ as seen in ARPES studies \cite{Yi2011}.

Inelastic neutron scattering studies of magnetic excitations in iron pnictides have been carried out extensively \cite{Zhao2008,Ewings2008,Zhao2009,Diallo2009,Ewings2011,Harriger2011,Lu2014}, and clearly reveal well defined spin wave excitations with energy scale $\sim$ 200 meV, persisting even above the N\'{e}el temperature \cite{Lu2014}, indicating that short range antiferromagnetic (AF) and ferromagnetic (F) order remain in the $a$ and $b$ directions, respectively, even above the disordering temperature for long-range magnetic order. This persistence of short range anisotropic magnetic order may account for the narrow nematic phase \cite{Kasahara2013,Zhou2013} above the N\'{e}el temperature where the ferro orbital order \cite{Yi2011} and structural distortion survive, as well as the temperature dependence of the measured anisotropies in $a$ and $b$ directions of magnetic excitations and resistivity \cite{Yi2011,Lu2014,Chu2010}.

Spin wave excitations in iron pnictides have been theoretically studied for multi-band models 
\cite{Kaneshita2010,Raghuvanshi2011,Knolle2011,Kovacic2015,Ghosh2015} in the random phase approximation (RPA), including the orbital matrix components of the spin wave spectral weight \cite{Kovacic2015}. The magnitude of the intra-orbital Coulomb interaction term considered in these investigations typically lie in the intermediate coupling range ($U \sim 1-2$ eV), resulting in moderately well developed local moments in the $(\pi,0)$ magnetic state, which has interesting implications on the multi-orbital quantum antiferromagnetism in these compounds. In this paper we will study the effective spin couplings generated by particle-hole exchange, their orbital contributions, and quantum corrections to sublattice magnetization in the $(\pi,0)$ AF state of a realistic three-orbital model which yields Fermi surface structure, spin wave excitations, and ferro orbital ordering in quantitative agreement with experiments. 

One important feature of the measured spin wave dispersion is that the spin wave energy is maximum at the ferromagnetic zone boundary (FZB), slightly higher than at the antiferromagnetic zone boundary (AFZB). This feature is quite significant as the FZB spin wave energy provides a sensitive measure of the effective ferromagnetic (F) spin coupling in the $(\pi,0)$ state \cite{Ghosh2015}. It is only when F spin coupling is included that the FZB spin wave energy becomes finite, and even maximum over the entire Brillouin zone when it exceeds the AF spin coupling. Understanding the microscopic mechanism behind the origin of this strong F spin coupling and its interplay with the complex multi-orbital electronic structure as observed in iron pnictides should provide significant insight toward understanding the multi-orbital antiferromagnetism in these compounds.

The structure of this paper is as below. Following a brief account in Section II of the realistic three band model for iron pnictides in terms of the $xz$, $yz$, and $xy$ Fe 3d orbitals, the effective spin couplings are introduced in Section III in terms of the particle-hole propagator $[\chi^0]$ evaluated in the $(\pi,0)$ magnetic state. Evaluation of quantum corrections to sublattice magnetization is discussed in Section IV in terms of transverse spin correlations $\langle S_{i\mu}^- S_{i\mu}^+ \rangle$ and $\langle S_{i\mu}^+ S_{i\mu}^- \rangle$. Results for the calculated effective spin couplings, quantum corrections, and discussion of their variation with $xy$ orbital energy offset are presented in Section V, followed by conclusions in Section VI. 

\section{Three-orbital model and ($\pi,0$) magnetic state}

We consider a minimal three-orbital model \cite{Ghosh2015} involving $d_{xz}$, $d_{yz}$ and $d_{xy}$ Fe $3d$ orbitals. The tight binding Hamiltonian in the plane-wave basis is defined as:
\begin{equation}
H_0 = \sum_{{\bf k},\sigma,\mu,\nu} T^{\mu\nu} ({\bf k}) a_{{\bf k}\mu\sigma}^{\dagger} a_{{\bf k}\nu\sigma}
\label{tight}
\end{equation}
where
\begin{eqnarray}
T^{11} &=& - 2t_1\cos  k_x - 2t_2\cos  k_y - 4t_3 \cos  k_x 
\cos  k_y \label{eq:t11}\nonumber \\
T^{22} &=& - 2t_2\cos  k_x -2t_1\cos  k_y - 4t_3 \cos  k_x 
\cos  k_y \label{eq:t22}\nonumber \\
T^{33} &=& - 2t_5(\cos  k_x+\cos  k_y)  - 4t_6\cos  k_x\cos  k_y 
+ \varepsilon_{xy}  \label{eq:t33} \nonumber \\
T^{12} &=& T^{21} =- 4t_4\sin  k_x \sin  k_y \label{eq:t12}\nonumber \\
T^{13} &=& \bar{T}^{31} = - 2it_7\sin  k_x - 4it_8\sin  k_x \cos  k_y
\label{eq:t13} \nonumber \\
T^{23} &=& \bar{T}^{32}= - 2it_7\sin  k_y - 4it_8\sin  k_y \cos  k_x\;
\label{eq:t23}
\end{eqnarray}
are the tight-binding matrix elements in the unfolded Brillouin Zone ($-\pi \leq k_x,k_y \leq \pi$). Here, $t_1$ and $t_2$ are the intra-orbital hoppings for $xz$ ($yz$) along $x$ ($y$) and $y$ ($x$) directions, respectively, $t_3$ and $t_4$ are the intra and inter-orbital hoppings along diagonal direction for $xz$ and $yz$, $t_5$ and $t_6$ are intra-orbital NN and NNN hoppings for $xy$, while $t_7$ and $t_8$ the NN and NNN hybridization between $xy$ and $xz/yz$. Finally, $\varepsilon_{ xy}$ is the energy difference between the $xy$ and degenerate $xz/yz$ orbitals.

\begin{figure}
\vspace*{0mm}
\hspace*{0mm}
\psfig{figure=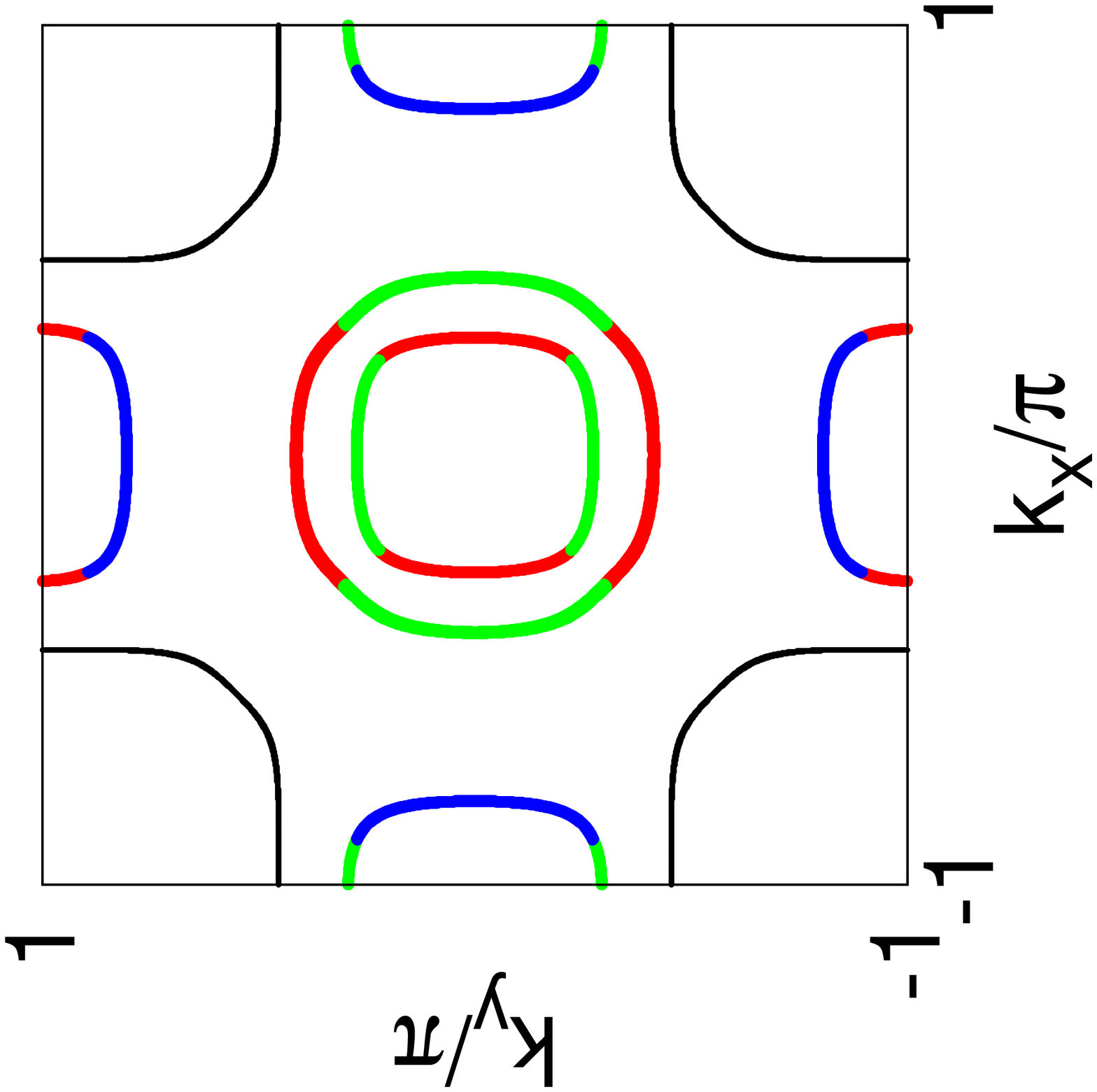,width=80mm}
\psfig{figure=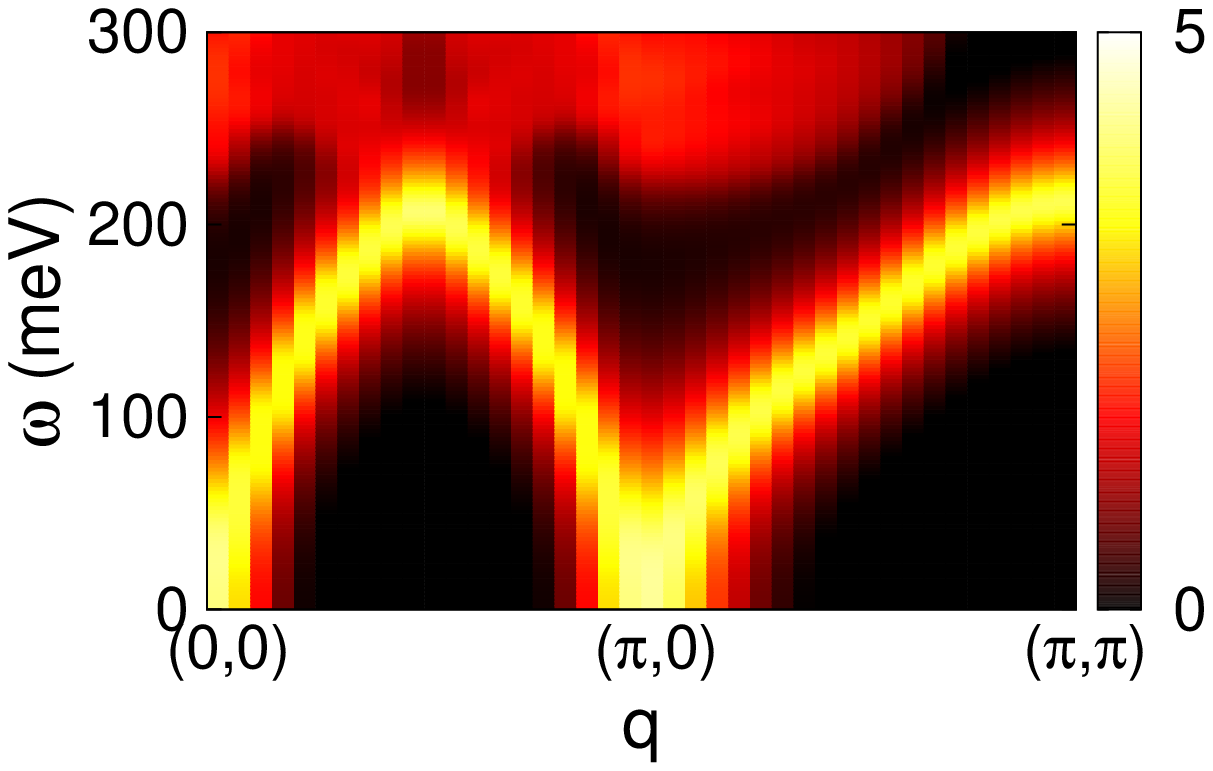,width=80mm}
\caption{(a) Fermi surface in the unfolded BZ for the three-orbital model \cite{Ghosh2015} with hopping parameters as given in Table \ref{hoppings}. The main orbital contributions are shown as: $d_{xz}$ (red), $d_{yz}$ (green), and $d_{xy}$ (blue). (b) The spin wave spectral function in the ($\pi,0$) SDW state for the three-orbital model at half filling.}
\end{figure}


\begin{table}
\caption{Values of the hopping parameters in the three-orbital model (in eV)} 
\begin{center}
\addtolength{\tabcolsep}{5pt}
 \begin{tabular}{c c c c c c c c c}\hline 
$t_1$ & $t_2$ & $t_3$ & $t_4$ & $t_5$ & $t_6$ & $t_7$ & $t_8$ \\ \hline
 $0.1$ & $0.32$ & $-0.29$ & $-0.06$ & $-0.3$ & $-0.16$ & $-0.15$ & $-0.02$ \\ \hline 
 \end{tabular}
\end{center}
 \label{hoppings}
\end{table}

The Fermi surface for the three-orbital Hamiltonian (\ref{tight}) for hopping parameter values given in Table 1 is shown in Figure 1(a) corresponding to near half filling. There are two circular hole pockets around the center and elliptical electron pockets around ($\pm \pi,0$) and ($0,\pm \pi$) in the unfolded BZ. The two hole pockets involve primarily the $xz$ and $yz$ orbitals, while the electron pockets centered at ($\pm \pi,0$) [($0,\pm \pi$)] arise mainly from the hybridization of the $xy$ and $yz$ [$xz$] orbitals. All of these features are in good agreement with results from DFT calculations and ARPES experiments. 

We now consider the ($\pi,0$) ordered magnetic (SDW) state of this model. The various electron-electron interaction terms  included are:

\begin{eqnarray}
H_I &=& U \sum_{i,\mu} n_{i\mu\uparrow} n_{i\mu\downarrow} + (U' - J/2) 
\sum_{i,\mu,\nu}^{\mu<\nu} n_{i\mu} n_{i\nu} - 2 J \sum_{i,\mu,\nu}^{\mu<\nu} 
{\bf {S_{i\mu}}} \cdot {\bf {S_{i\nu}}} \nonumber \\
&+& J' \sum_{i,\mu,\nu}^{\mu<\nu}
(a_{i\mu\uparrow}^{\dagger} a_{i\mu\downarrow}^{\dagger} a_{i\nu\downarrow} a_{i\nu\uparrow} + \rm{H.c.}),
\label{interaction}
\end{eqnarray}
where ${\bf S}_{i\mu}$ ($n_{i\mu}$) refer to the local spin (charge) density operators for orbital $\mu$. The first and second terms are the intra-orbital and inter-orbital Coulomb interactions respectively, the third term is the Hund's rule coupling and the fourth term the ``pair-hopping'' term. In the following, we will consider $U = 1.2$ eV and $J \approx U/4$. This interaction strength corresponds to the intermediate coupling regime, which is in accord with recent DFT + DMFT study of magnetism in iron pnictides \cite{Yin2011}.

Extending the two-sublattice basis approach for the SDW state in a single-band model \cite{Raghuvanshi2011} to a composite three-orbital, two-sublattice basis, the Hartree-Fock (HF) level Hamiltonian matrix in this composite basis (A$xz$ A$yz$ A$xy$ B$xz$ B$yz$ B$xy$) is obtained as:

\begin{footnotesize}
\begin{eqnarray}
H_{\rm HF}^{\sigma} ({\bf k}) =  
\left [ \begin{array}{cccccc} -\sigma
\Delta_{xz} + \varepsilon_{\bf k}^{2y}   & 0 & 0 & \varepsilon_{\bf k}^{1x}
+ \varepsilon_{\bf k}^{3} & \varepsilon_{\bf k}^{4} & \varepsilon_{\bf k}^{7x}
+ \varepsilon_{\bf k}^{8,1}
\\ 
0 & -\sigma \Delta_{yz} + \varepsilon_{\bf k}^{1y} & \varepsilon_{\bf k}^{7y}  &
\varepsilon_{\bf k}^{4} & \varepsilon_{\bf k}^{2x} + \varepsilon_{\bf k}^{3}  &
\varepsilon_{\bf k}^{8,2} \\ 
0 & -\varepsilon_{\bf k}^{7y} & - \sigma \Delta_{xy} + \varepsilon_{\bf
k}^{5y} + \varepsilon_{xy} & - \varepsilon_{\bf k}^{7x}- \varepsilon_{\bf
k}^{8,1} & -\varepsilon_{\bf k}^{8,2} & \varepsilon_{\bf k}^{5x} +
\varepsilon_{\bf k}^{6} \\ 
\varepsilon_{\bf k}^{1x} + \varepsilon_{\bf k}^{3} & \varepsilon_{\bf k}^{4} &
\varepsilon_{\bf k}^{7x} + \varepsilon_{\bf k}^{8,1} & \sigma \Delta_{xz} +
\varepsilon_{\bf k}^{2y} & 0 & 0 \\
\varepsilon_{\bf k}^{4} & \varepsilon_{\bf k}^{2x} + \varepsilon_{\bf k}^{3} &
\varepsilon_{\bf k}^{8,2} & 0 & \sigma \Delta_{yz} + \varepsilon_{\bf k}^{1y} &
\varepsilon_{\bf k}^{7y} \\
- \varepsilon_{\bf k}^{7x} - \varepsilon_{\bf k}^{8,1} & -\varepsilon_{\bf
k}^{8,2} & \varepsilon_{\bf k}^{5x} + \varepsilon_{\bf k}^{6} & 0 &
- \varepsilon_{\bf k}^{7y} & \sigma
\Delta_{xy} + \varepsilon_{\bf k}^{5y} + \varepsilon_{xy}  \\
\end{array}
\right ] \nonumber \\ 
\label{Hamiltonian}
\end{eqnarray}
\end{footnotesize}
for spin $\sigma$, in terms of the band energies corresponding to hopping terms along different directions:
\begin{eqnarray}
\varepsilon_{\bf k}^{1x} &=& -2 t_1 \cos k_x  \;\;\;\;\;\;
\varepsilon_{\bf k}^{1y} = -2 t_1 \cos k_y  \nonumber \\
\varepsilon_{\bf k}^{2x} &=& -2 t_2 \cos k_x  \;\;\;\;\;\; 
\varepsilon_{\bf k}^{2y} = -2 t_2 \cos k_y  \nonumber \\
\varepsilon_{\bf k}^{5x} &=& -2 t_5 \cos k_x  \;\;\;\;\;\; 
\varepsilon_{\bf k}^{5y} = -2 t_5 \cos k_y  \nonumber \\
\varepsilon_{\bf k}^{3} &=& -4 t_3 \cos k_x \cos k_y  \;\;\;\;\;\;
\varepsilon_{\bf k}^{4} = -4 t_4 \sin k_x \sin k_y \nonumber \\ 
\varepsilon_{\bf k}^{6} &=& -4 t_6 \cos k_x \cos k_y \nonumber \\
\varepsilon_{\bf k}^{7x} &=& -2i t_7 \sin k_x  \;\;\;\;\;\;
\varepsilon_{\bf k}^{7y} = -2i t_7 \sin k_y  \nonumber \\
\varepsilon_{\bf k}^{8,1} &=& -4i t_8 \sin k_x \cos k_y  \;\;\;\;\;\;
\varepsilon_{\bf k}^{8,2} = -4i t_8 \cos k_x \sin k_y \nonumber \\ 
\end{eqnarray}
and the self-consistent exchange fields defined as $2\Delta_\mu = U m_{\mu} + J\sum_{\nu \neq \mu}m_{\nu}$ in terms of sublattice magnetization $m_{\mu}$ for orbital $\mu$. 

The calculated spin wave spectral function \cite{Ghosh2015} in the SDW state for the three-orbital model is shown in Fig. 1(b). Evidently, spin wave excitations are highly dispersive, and do not decay into the particle-hole continuum. The energy scale of spin excitations is $\sim$ 200 meV with a well-defined maximum at the ferromagnetic zone boundary [${\bf q}=(0,\pi)$]. These features of spin wave excitations are in excellent agreement with results from inelastic neutron scattering measurements, confirming the realistic nature of the three-orbital model.

\section{Effective spin couplings}
In the $(\pi,0)$ magnetic state of the multi-band interacting electron model, the transverse spin fluctuation (spin wave) propagator in the random phase approximation (RPA) can be expressed in a symmetrised form:
\begin{eqnarray}
[\chi^{-+}({\bf q},\omega)] & = & \frac{ [\chi^0({\bf q},\omega)] } 
{ {\bf 1} - [U][\chi^0({\bf q},\omega)] } \nonumber \\
& = & \frac{\bf 1} { [U] - [U][\chi^0({\bf q},\omega)][U] } - \frac{\bf 1} { [U] }
\end{eqnarray}
where $[\chi^0]$ is the bare particle-hole propagator in the orbital-sublattice basis evaluated by integrating out the fermionic degrees of freedom, $[U]$ is the local (on-site) interaction matrix with intra-orbital terms $U$ and inter-orbital (Hund's rule coupling) terms $J$, and the matrix $[U][\chi^0({\bf q},\omega)][U]$ is Hermitian. 

Broadly, the important features of spin wave excitations in the spontaneously-broken-symmetry $(\pi,0)$ magnetic state of the three-band model are: (i) presence of zero-energy Goldstone modes related to continuous spin rotation symmetry of the interacting electron Hamiltonian and (ii) coupling between the three orbitals due to orbital hybridization in the tight-binding model. We therefore consider mapping to an effective spin model:
\begin{equation}
{\cal H} = \sum_{\langle ij \rangle, \mu\nu} J_{ij}^{\mu\nu} {\bf S}_{i\mu} . {\bf S}_{j\nu} 
\end{equation}
in terms of spin-1/2 operators ${\bf S}_{i\mu}$ corresponding to the three electronic orbitals $\mu= xz,yz,xy$ which allows for effective spin interactions (couplings) between different orbitals while preserving the spin rotation symmetry. 

Since positive spin wave energies are associated with increase in spin interaction energy corresponding to specific spin twisting modes, the effective spin couplings are inherently present in the RPA level spin propagator of Eq. (6). Based on earlier spin wave studies in interacting electron models, as briefly discussed below, we will consider the correspondence:

\begin{equation}
-\frac{J_{ij}^{\mu\nu}}{2} = \sum_{\bf q} \sum_{\mu' \nu'} \; [U]^{\mu\mu'} \; [\chi^0({\bf q})]^{\mu'\nu'} _{ss'} \; [U]^{\nu'\nu} \; e^{i {\bf q}.({\bf r}_i - {\bf r}_j)}
\end{equation}
where indices $s,s'$ correspond to the sublattices (A/B) which sites $i$ and $j$ belong to, and we have set $\omega=0$ in the bare particle-hole propagator $[\chi^0({\bf q},\omega)]$. The above form clearly suggests exchange of the particle-hole propagator $[\chi^0]$ as the origin of the effective inter-site spin couplings in interacting electron models with purely local interactions of the form $-U_{\mu\mu'} {\bf S}_{i\mu} . {\bf S}_{i\mu'}$.
In the following, we will focus on the first neighbor spin couplings in the $a$ (AF) and $b$ (F) directions (1a and 1b) and the second neighbor coupling along the diagonal direction (2), evaluated by considering the corresponding lattice connectivity vectors ${\bf \delta} = {\bf r}_i - {\bf r}_j$ in Eq. (8). 

In the $(\pi,\pi)$ AF state of the single-band Hubbard model at half filling, extensively studied in the context of quantum antiferromagnetism in cuprates, the above prescription yields AF spin coupling $4t^2/U$ in the strong coupling limit \cite{Singh1990}. In the ferromagnetic state of the Ferromagnetic Kondo Lattice Model (FKLM), which has been extensively studied for understanding metallic ferromagnetism in doped manganites, the calculated spin wave dispersion in the strong coupling (double exchange) limit $(J_{\rm H} /t \rightarrow \infty)$ corresponds exactly to that for the Quantum Heisenberg Ferromagnet with NN spin coupling given by $J_{\rm H}^2 \chi^0 ({\bf q})$ \cite{FKLM2008}. 

Perhaps most relevant is that in the $(\pi,0)$ magnetic state of the half filled $t-t'-U$ model, this correspondence exactly yields the AF first and second neighbor spin couplings $J_1 = 4t^2/U$ and $J_2 = 4t^{' 2}/U$ in the strong coupling limit, and also the effective F spin couplings of opposite sign for the doped $t-t'-U$ model where the F spin couplings are generated due to particle-hole exchange in the partially filled band as in metallic ferromagnets \cite{Ghosh2015}. Finally, in the limit of vanishing electron interaction strength (free electron limit), the above prescription reduces exactly to the RKKY interaction, and thus allows for continuous interpolation between the weak and strong coupling limits. It should be noted that dynamical effects on effective spin couplings (through $\omega$ dependence in $[\chi^{(0)} ({\bf q},\omega)]$ have been neglected, but should not affect the results qualitatively in the intermediate coupling regime. 

\section{Quantum corrections to sublattice magnetization}

Towards understanding the quantum antiferromagnetism in cuprate antiferromagnets, quantum spin fluctuations have been studied intensively in view of their important role in diverse macroscopic properties such as existence of long-range order, quantum corrections to sublattice magnetization, perpendicular susceptibility, spin wave velocity, ground state energy, and spin correlations \cite{manousakis,quantum}. In terms of the half-filled Hubbard model representation for the AF ground state, quantum spin fluctuations reduce the sublattice magnetization to nearly $60\%$ of the classical (HF) value in two dimensions in the strong-coupling limit $(U/t\rightarrow \infty)$. The sublattice magnetization quantum corrections for the individual orbitals in the $(\pi,0)$ state of the three band model are therefore of interest. We will use the approach in terms of transverse spin correlations which is valid in the full range of interaction strength for the Hubbard model and interpolates properly to the strong coupling limit \cite{trans}. Extending this approach to the multi-orbital antiferromagnet, the corrected sublattice magnetization for orbital $\mu$ is obtained from: 
\begin{equation}
m_\mu = m_\mu ^{\rm HF} - \delta m_\mu ^{\rm SF}
\end{equation} 
where the first-order, quantum spin-fluctuation corrections:
\begin{equation}
\delta m_\mu ^{\rm SF} = 
\frac{\langle S_{i\mu} ^+ S_{i\mu} ^- \rangle + \langle S_{i\mu} ^- S_{i\mu} ^+ \rangle } 
{\langle S_{i\mu} ^+ S_{i\mu} ^- \rangle - \langle S_{i\mu} ^- S_{i\mu} ^+ \rangle } -1 .
\end{equation}

The transverse spin correlations above (equal-time, same-site) are evaluated from the retarded part of the RPA level transverse spin fluctuation propagator:
\begin{eqnarray}
\langle S^- _{i\mu} (t) S^+ _{i\mu} (t'\rightarrow t^-)\rangle &=& 
\sum_{\bf q} \int_0 ^\infty -\frac{d\omega}{\pi} {\rm Im} [\chi^{-+}({\bf q},\omega)]_{\mu\mu} ^{AA} \nonumber \\
\langle S^+ _{i\mu} (t) S^- _{i\mu} (t'\rightarrow t^-)\rangle &=& 
\sum_{\bf q} \int_0 ^\infty -\frac{d\omega}{\pi} {\rm Im} [\chi^{+-}({\bf q},\omega)]_{\mu\mu} ^{AA} \nonumber \\
&=& \sum_{\bf q} \int_0 ^\infty -\frac{d\omega}{\pi} {\rm Im} [\chi^{-+}({\bf q},\omega)]_{\mu\mu} ^{BB} 
\end{eqnarray}
for lattice site $i$ on A sublattice, using the spin-sublattice symmetry in the AF state which relates correlations on A and B sublattices via $\langle S_{i\mu} ^+ S_{i\mu} ^- \rangle_A = \langle S_{i\mu} ^- S_{i\mu} ^+ \rangle_B$.

\section{Results and discussion}

\subsection{Effective spin couplings}
The effective spin couplings $J_{\bf \delta}^{\mu\nu}$ were calculated from Eq. (8), with the ${\bf q}$ summation over the two dimensional Brillouin Zone performed for grids upto $60 \times 60$ to ensure no significant variation. The results are shown in Table 2 for the first and second neighbors indicated by ${\bf \delta} \equiv 1a,1b,2$, and for the reference case $\epsilon_{xy}$=0.8 eV as considered for the spin wave plot of Fig. 1(b). Among the diagonal terms, the F (negative) spin coupling ($1b$) is maximum for the $xy$ orbital, whereas AF spin couplings are strong for $xy$ and $yz$ orbitals ($1a$) and for $xz$ and $yz$ orbitals ($2$). The dominant off-diagonal term is the AF spin coupling ($1a$) involving the $xz$ and $xy$ orbitals. Interestingly, this off-diagonal term is the maximum AF spin coupling, highlighting the multi-orbital character of quantum antiferromagnetism and the role of strong orbital hybridization in the three band model.

As all three orbitals follow identical $(\pi,0)$ magnetic ordering, it is useful to consider the total (orbital summed) spin couplings 
\begin{equation}
J_{\bf \delta} =  \sum_{\mu,\nu} J_{\bf \delta} ^{\mu\nu}  
\end{equation}
which yield the effective couplings as considered in phenomenological spin models to describe the spin wave dispersion in iron pnictides \cite{Zhao2009}. The most significant feature of the total spin couplings, given in Table 3, is the  ferromagnetic (negative) first neighbor spin coupling ($1b$), which is in agreement with Ref. \cite{Zhao2009} where it was shown that spin wave dispersion throughout the BZ and the maximum at ($\pi,\pi$) can be explained by a suitably parameterized Heisenberg Hamiltonian with an effective ferromagnetic exchange interaction ($J_{1b}<0$) in the $b$ direction. The present work provides the microscopic origin of this ferromagnetic interaction as due to the usual particle-hole exchange process in an itinerant-electron model, with the $xy$ orbital being mainly responsible for the strong F spin coupling generated.

We now investigate the effect of orbital order $n_{xz} - n_{xy}$ (controlled by the offset energy $\varepsilon_{xy}$) on spin wave energies and effective spin couplings. Fig. 2 shows the variation of (a) the electronic densities $(n_\mu)$ for the three orbitals and (b) the spin wave energies at the ferromagnetic ${\bf q} = (0,\pi)$ and antiferromagnetic $(\pi/2,0)$ zone boundaries. While $n_{yz}$ remains fixed at 1 (half filling), there is a significant transfer of electronic density from $xy$ to $xz$ orbital with increasing $xy$ orbital energy offset. The strong enhancement in the FZB spin wave energy with the depletion in $xy$ orbital electronic density from 1 is due to the corresponding enhancement in the F spin coupling contribution of the $xy$ orbital, as discussed below. 

\begin{table}[!]
\centering
\label{Table-1}
\begin{tabular}{c | c c c}
 $J^{\mu\nu}_{1a}$ (meV) & $xz$ & $yz$ & $xy$ \\
 \hline
 $xz$ & 6.22 & 15.54 & 40.33 \\ 
 
 $yz$ & 15.54 & 32.32 & 21.06 \\
 
 $xy$ & 40.33 & 21.06 & 34.09 \\ 
\end{tabular}
\\
\begin{tabular}{c | c c c}
 $J^{\mu\nu}_{1b}$ (meV) & $xz$ & $yz$ & $xy$ \\
 \hline
 $xz$ & 17.06 & 0.53 & -5.75 \\ 
 
 $yz$ & 0.53 & -17.17 & -2.62 \\
 
 $xy$ & -5.75 & -2.62 & -48.74 \\ 
\end{tabular}
\\
\begin{tabular}{c | c c c}
 $J^{\mu\nu}_{2}$ (meV) & $xz$ & $yz$ & $xy$ \\
 \hline
 $xz$ & 28.86 & 15.38 & -5.16 \\ 
 
 $yz$ & 15.38 & 31.66 & 0.67 \\
 
 $xy$ & -5.16 & 0.67 & -13.44 \\ 

\end{tabular}
\caption{Effective spin couplings $J^{\mu\nu}_{\bf \delta}$ for first and second neighbors, evaluated from Eq. (8) for the reference case $\epsilon_{xy} = 0.8$ eV.}
\end{table}

\begin{table}[!]
\centering
\label{table:2}
\begin{tabular}{c | c c c}
 ${\bf \delta}$ & $1a$ & $1b$ & $2$ \\ 
 \hline
$J_{\delta}$ (meV) & 226.48 & -64.52 & 68.86 \\  
 
\end{tabular}
\caption{The total (orbital summed) effective spin couplings for first and second neighbors.}
\end{table}

\begin{figure}
\vspace*{0mm}
\hspace*{0mm}
\psfig{figure=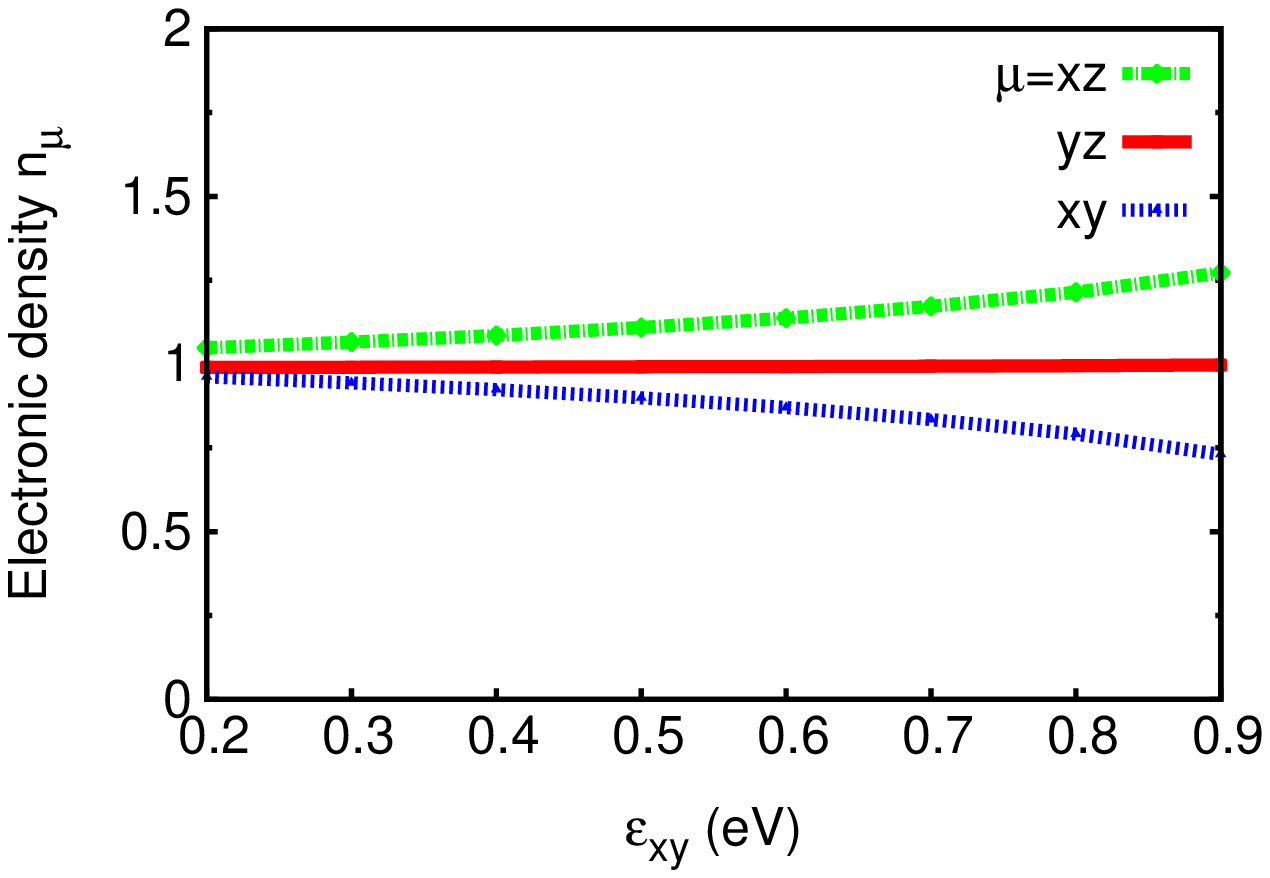,width=80mm}
\psfig{figure=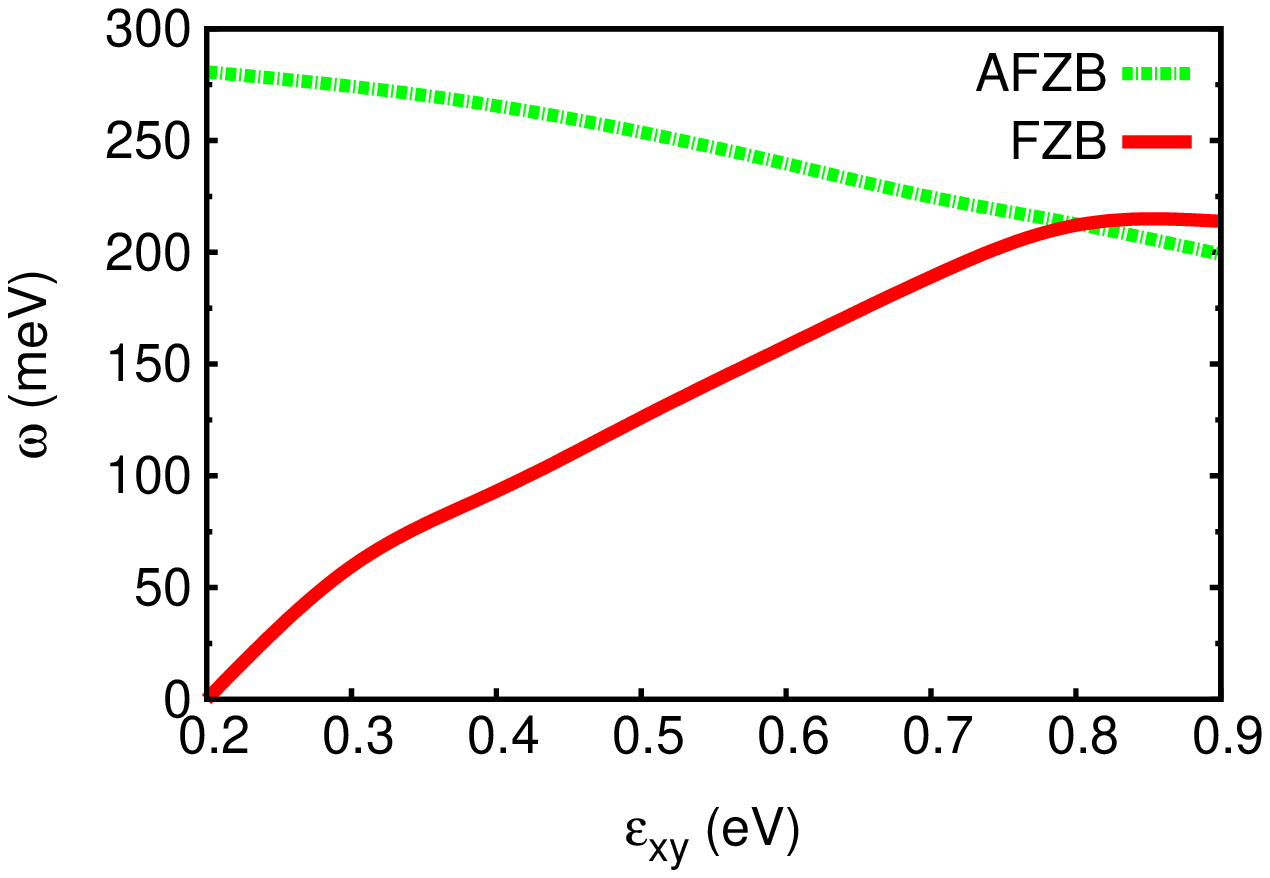,width=80mm}
\caption{Variation of (a) electronic density $n_\mu$ for the three orbitals and (b) spin wave energy at the F and AF zone boundaries with the $xy$ orbital energy offset $\varepsilon_{xy}$.}
\label{effective_couplings}
\end{figure}


Fig. 3(a) shows the variation (again with $\varepsilon_{xy}$) of the total (orbital summed) spin couplings for the first neighbors ($1a,1b$) in the $a$ and $b$ directions and the second neighbor (2) along the diagonal direction. For higher values of the $xy$ orbital energy offset, the effective spin coupling $J_{1b}$ is strongly negative, clearly highlighting the strong F spin generated by the particle-hole process mainly in the partially filled $xy$ band. With decreasing $xy$ orbital offset energy, as the $xy$ band approaches half filling, the F component of the spin coupling generated by particle-hole exchange decreases substantially, resulting in the net spin coupling changing sign from negative to positive. 

\begin{figure}
\vspace*{0mm}
\hspace*{0mm}
\psfig{figure=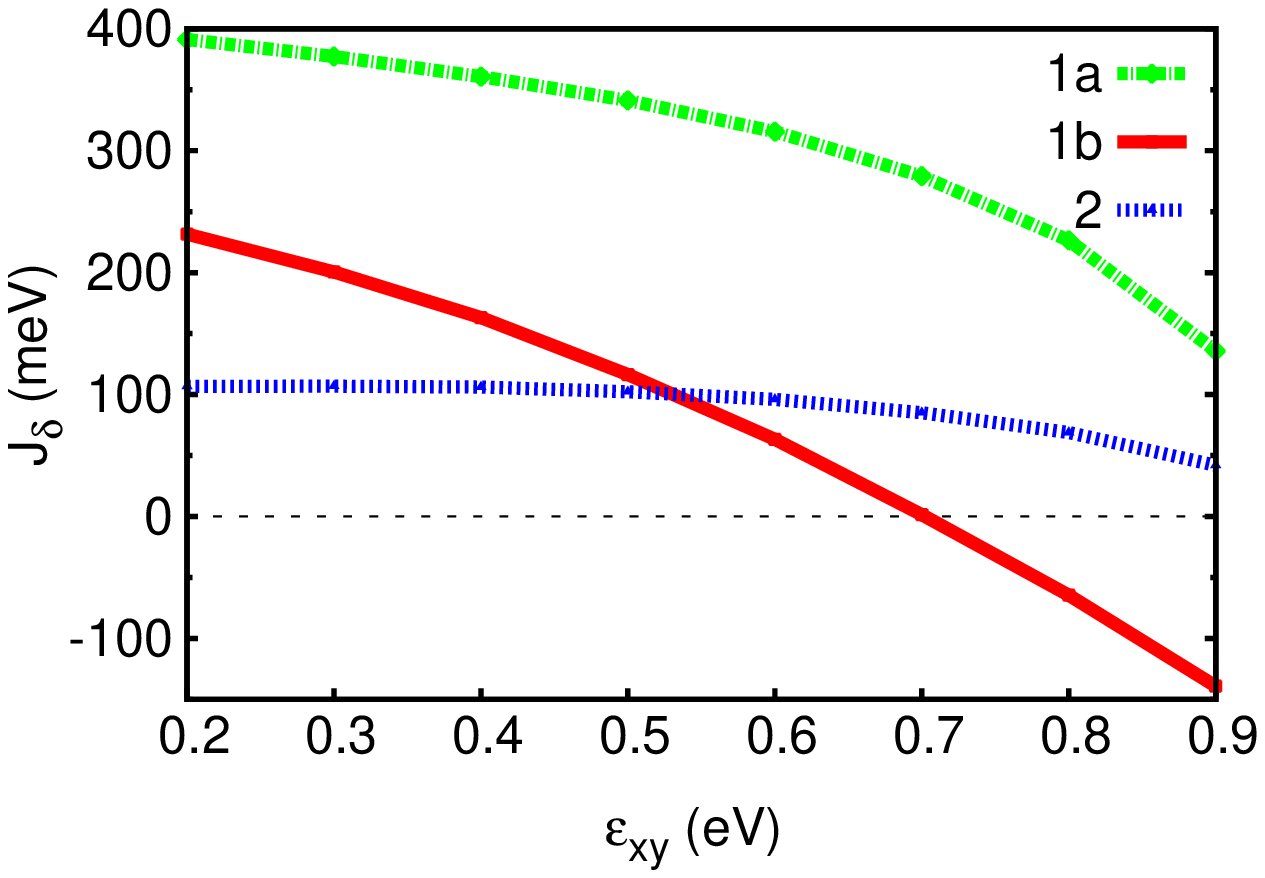,width=80mm}
\psfig{figure=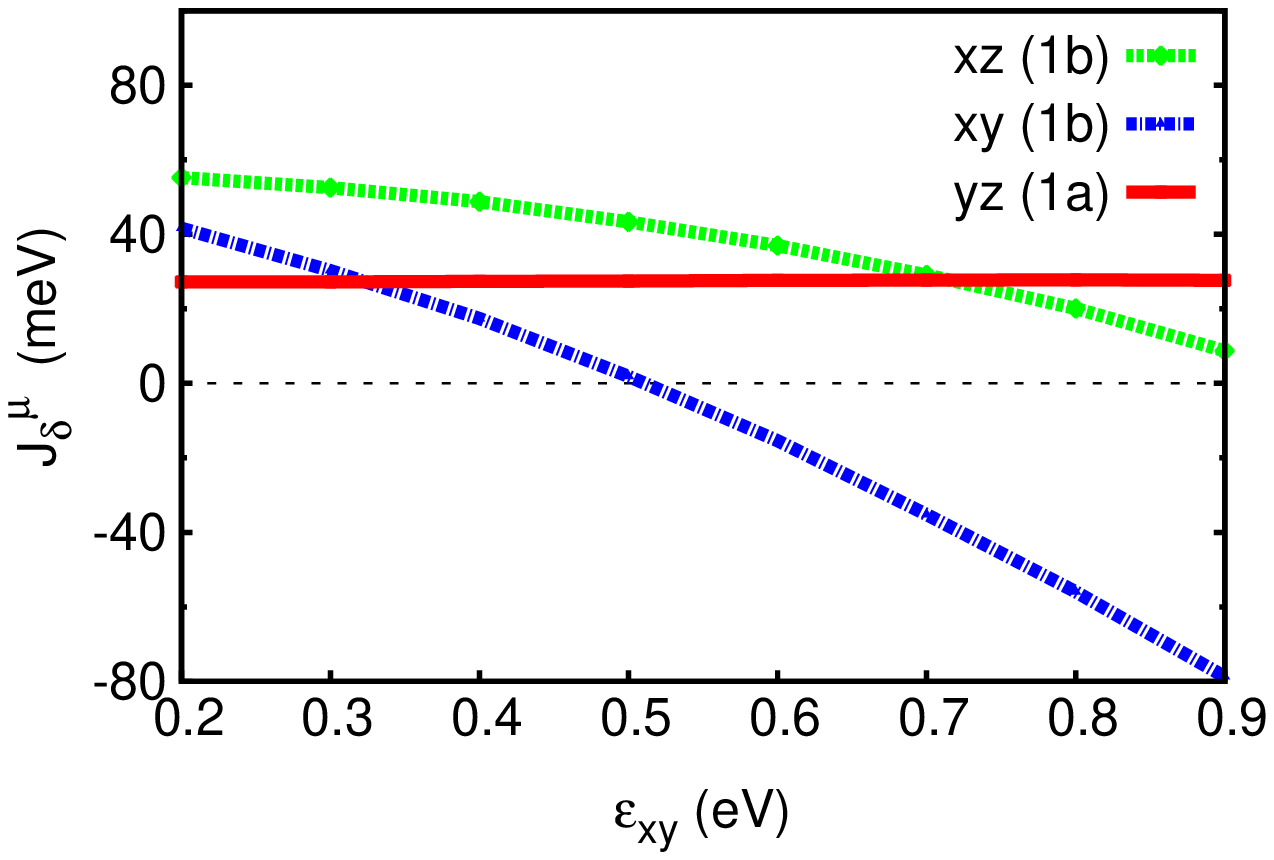,width=80mm}
\caption{Variation of the (a) total (orbital summed) spin couplings $J_{\bf \delta}$ and (b) the individual orbital contributions $J_{\bf \delta} ^{\mu}$ to the effective spin couplings for the first and second neighbors with the $xy$ orbital energy offset $\varepsilon_{xy}$.}
\label{orbital_couplings}
\end{figure}


Due to orbital hybridization in the three band tight binding model (1), the effective spin couplings considered earlier include contributions from the orbital off-diagonal parts $[\chi^0]^{\mu \ne \nu}$ of the particle-hole propagator as well. We therefore consider the individual orbital contributions to the effective spin couplings:
\begin{equation}
-\frac{ J_{\bf \delta} ^{\mu} }{2} = U^2  \sum_{\bf q} \; [\chi^0({\bf q})]^{\mu\mu} _{ss'} \; 
e^{i {\bf q}.({\bf r}_i - {\bf r}_j)}
\end{equation}
which are shown in Fig. 3(b) for the three orbitals $\mu=xz,yz,xy$. The $xy$ orbital yields strong F spin coupling $(1b)$ for higher values of $xy$ orbital offset energy, which sharply decreases in magnitude, even changing sign to AF spin coupling, with decreasing $xy$ orbital offset energy. This reduction is evidently related to the $xy$ orbital approaching half filling, which strongly suppresses the F spin coupling generated by the particle-hole exchange. 

We now address the origin of the strong F spin coupling generated in the $xy$ orbital sector as seen in Fig. 3(b). Ferro spin couplings are generated generically through exchange of the particle-hole propagator, as in metallic ferromagnets with partially filled bands, and are especially strong when the Fermi energy lies near a van Hove peak in the DOS, which enhances the favorable delocalization contribution and suppresses the unfavorable exchange contribution \cite{Singh2006}. At first sight, it may seem that the electronic structure of the gapped $(\pi,0)$ magnetic state has little in common with metallic ferromagnets where the Fermi energy lies within the band. However, the $xy$ orbital sector is indeed only partially filled [Fig. 2(a)] in the gapped $(\pi,0)$ state due to the strong hybridization with $xz/yz$ orbitals, and we attribute the generation of F spin coupling in the $b$ direction to particle-hole processes in this partially filled $xy$ sector.

Indeed, the F (negative) spin coupling is maximum at the upper end of the scale [Fig. 3(b)] where the $xy$ electronic density is maximally depleted $(n_{xy} \approx 0.75)$ and the energy gap for particle-hole excitations is also minimum, which together result in strong enhancement of the F spin coupling generated by particle-hole exchange. With decreasing $xy$ orbital energy offset, as the $xy$ orbital approaches half filling ($n_{xy} \approx 1$), the particle-hole-exchange mediated F component of the spin coupling decreases substantially, resulting in the net spin coupling changing sign from negative to positive. 

As also seen from Fig. 3(b), the AF spin coupling $(1a)$ for the half-filled $yz$ orbital shows no variation with $xy$ orbital energy offset. For the $xz$ orbital, the strongly AF spin coupling $(1b)$ near half filling (due to large hopping term $t_2$) decreases substantially with $xy$ orbital offset energy as $xz$ band filling increases beyond 1 [Fig. 2(a)], indicating F contribution from the additional states transferred from the $xy$ band. This reduced AF spin coupling in the ferromagnetic ($b$) direction also contributes to the stabilization of the $(\pi,0)$ state by suppressing magnetic frustration. 

\begin{figure}
\vspace*{0mm}
\hspace*{0mm}
\psfig{figure=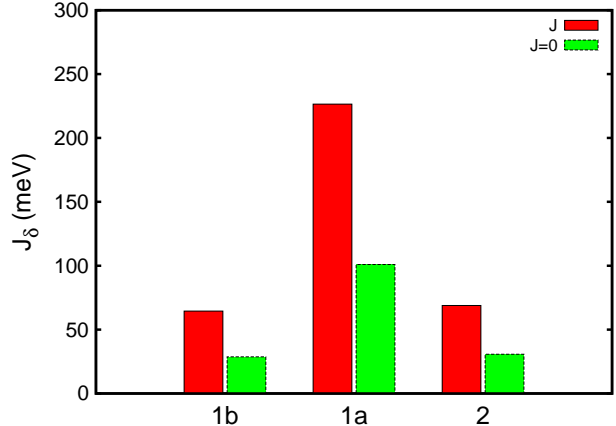,width=80mm}
\caption{Comparison of the effective spin couplings $J_{\bf \delta}$ with and without the Hund's rule coupling term $J$, highlighting the significant role of $J$ in stabilizing the ($\pi,0$) state.}
\label{histogram}
\end{figure}


Each spin coupling term (8) involves altogether nine terms corresponding to $\mu'\nu'=xz,yz,xy$, out of which eight terms involve the Hund's rule coupling term $J$, indicating the importance of $J$ in the effective spin couplings. Fig. 4 shows the total (orbital summed) spin couplings $J_{\bf \delta} = \sum_{\mu\nu} J^{\mu \nu} _{\bf \delta}$ evaluated with and without $J$ for the three neighbors. The significant reduction (to nearly half in all three cases) in the total spin couplings when $J$ is set to zero highlights the importance of the Hund's rule coupling term in the effective spin couplings and therefore on the overall stabilization of the $(\pi,0)$ magnetic state.  

\subsection{Quantum corrections to sublattice magnetization}

For evaluating the quantum corrections to sublattice magnetization, the ${\bf q}$ summation over the BZ in Eq. (11) was performed over a $60 \times 60$ grid as earlier. The $\omega$ integral was performed by evaluating the transverse spin fluctuation spectral function for 1000 $\omega$ points extending upto 4 eV to ensure that optical modes and particle-hole excitations up to highest energies are included. Besides the Goldstone mode, optical modes are also generally present for multi-orbital magnetic systems. The calculated transverse spin correlations and quantum corrections to sublattice magnetization are given in Table 4 for the three orbitals, showing maximum and minimum quantum corrections for the $xz$ and $yz$ orbitals, respectively. 

It should be noted that the spin identities following from the spin commutation relations:
\begin{equation}
\langle [ S_{i\mu} ^+ , S_{i\mu} ^- ] \rangle = \langle 2S_{i\mu} ^z \rangle
\end{equation}
between the RPA level transverse spin correlations and the HF level magnetizations provide a stringent check on the numerical accuracy, and are obeyed to high degree as seen from Table 4. 

\begin{table}[!]
\centering
\label{table:4}
\begin{tabular}{c | c | c | c | c | c}
$\mu$ & $\langle S_{i\mu}^+ S_{i\mu}^- \rangle$ & $\langle S_{i\mu}^- S_{i\mu}^+ \rangle$ & $\langle [S_{i\mu}^+ ,S_{i\mu}^-]\rangle$ & $\langle 2S_{i\mu}^z \rangle$ & $\delta m_{i\mu}$ \\ 
\hline
$xz$ & 0.64 & 0.088 & 0.55 & 0.57 & 0.32 \\
$yz$ & 0.79 & 0.054 & 0.73 & 0.75 & 0.15 \\
$xy$ & 0.70 & 0.081 & 0.62 & 0.63 & 0.26 \\
\end{tabular}
\caption{Transverse spin correlations and sublattice magnetization quantum corrections for the three orbitals.}
\end{table}

With the total (orbital summed) quantum correction to sublattice magnetization $\sum_\mu \delta m_\mu = 0.73$, and the total HF level sublattice magnetization $\sum_\mu m^{\rm HF}_\mu = \sum_\mu \langle 2S^z _\mu \rangle  = 1.95$, the orbital-averaged reduction in sublattice magnetization due to quantum spin fluctuation induced quantum corrections in the $(\pi,0)$ AF state of the three band model is quite significant at about $37\%$.






\section{Conclusions}

Originating from particle-hole exchange, effective spin couplings $J_{\bf \delta}^{\mu\nu}$ for first and second neighbors were evaluated in the $(\pi,0)$ magnetic state of a realistic three band interacting electron model for iron pnictides involving $xz,yz,xy$ Fe 3d orbitals. Variation of these spin couplings with the $xy$ orbital energy offset provides valuable insight into the multi-orbital quantum antiferromagnetism in these compounds. The $xy$ orbital was found to be mainly responsible for the generation of strong F spin coupling between first neighbors in the $b$ direction, which is critically required to fully account for the spin wave dispersion measured from inelastic neutron scattering experiments. The F spin coupling is strongly suppressed with decreasing orbital order as the $xy$ band approaches half filling, and is ascribed, as in metallic ferromagnets, to particle-hole exchange in the partially filled $xy$ band, which provides the microscopic basis for the negative (ferromagnetic) exchange interaction ($J_{1b}<0$) as considered in phenomenological spin models.

Significantly, the strongest AF spin coupling between first neighbors in the $a$ direction lies in the orbital off diagonal sector involving the $xz$ and $xy$ orbitals, highlighting the important role of strong orbital hybridization on effective spin couplings. While the AF spin coupling in $a$ direction for the half-filled $yz$ orbital was found to be constant with increasing $xy$ orbital energy offset, the frustrating AF spin coupling in $b$ direction for the $xz$ orbital was found to decrease, thus reducing the magnetic frustration substantially. The Hund's rule coupling term was found to contribute significantly to the effective spin couplings and therefore to the overall stabilization of the $(\pi,0)$ magnetic state.  

The first-order spin-fluctuation induced quantum corrections to sublattice magnetization were evaluated from the transverse spin correlations, and yield maximum reduction for the overfilled $xz$ orbital and minimum reduction for the half filled $yz$ orbital. The orbital-averaged reduction in sublattice magnetization due to quantum spin fluctuations in the $(\pi,0)$ AF state of the three band model was found to be significant at about $37\%$.

\section*{Acknowledgements}
SG and NR acknowledge financial support from the Council of Scientific and Industrial Research, India.

\end{document}